\documentclass[twocolumn]{aastex631}

\newcommand{\oi}{i_{0}}
\newcommand{\ogi}{(g-i)_{0}}

\newcommand{\mv}{\rm{M_{V}}}

\newcommand{\mg}{\rm{M_{g}}}
\newcommand{\mi}{\rm{M_{i}}}
\newcommand{\mm}{(m-M)_{0}}
\newcommand{\gmh}{\rm{[M/H]}}

\newcommand{\rc}{\rm{R_{c}}}
\newcommand{\rt}{\rm{R_{t}}}
\newcommand{\roc}{\rm{R_{c(out)}}}

\newcommand{\re}{\rm{R_{exp}}}
\newcommand{\rh}{\rm{R_{h}}}
\newcommand{\roh}{\rm{R_{h(out)}}}

\newcommand{\msun}{\rm{M}_{\odot}}
\newcommand{\ohmetal}{12+\log(\rm{O/H})}
\newcommand{\rb}{\rm{R_{B}}}
\newcommand{\tcr}{\rm{t_{cr}}}

\begin{document}

\title{The Progenitor of the Peculiar Galaxy NGC\,3077 \footnote{Based on data collected at Subaru Telescope, which is operated by the National Astronomical Observatory of Japan.}}

\author[0000-0002-7866-0514]{Sakurako Okamoto}
\affiliation{Subaru Telescope, National Astronomical Observatory of Japan,\\ 650 North A'ohoku Place, Hilo, HI 96720, U.S.A.}
\affiliation{National Astronomical Observatory of Japan, Osawa 2-21-1, Mitaka, Tokyo, 181-8588, JAPAN}
\affiliation{The Graduate University for Advanced Studies, Osawa 2-21-1, Mitaka, Tokyo 181-8588, Japan}

\author{Nobuo Arimoto}
\affiliation{National Astronomical Observatory of Japan, Osawa 2-21-1, Mitaka, Tokyo, 181-8588, JAPAN}
\affiliation{The Graduate University for Advanced Studies, Osawa 2-21-1, Mitaka, Tokyo 181-8588, Japan}
\affiliation{Astronomy Program, Department of Physics and Astronomy, Seoul National University, 599 Gwanak-ro, Gwanak-gu, Seoul, 151-742, Korea}

\author[0000-0001-7934-1278]{Annette M.N. Ferguson}
\affiliation{Institute for Astronomy, University of Edinburgh, Royal Observatory, Blackford Hill, Edinburgh, EH9 3HJ U.K.}

\author[0000-0002-2191-9038]{Mike J. Irwin}
\affiliation{Institute of Astronomy, University of Cambridge, Madingley Road, Cambridge CB3 0HA, U.K.}

\author[0000-0002-8566-0491]{Rokas \v{Z}emaitis}
\affiliation{Institute for Astronomy, University of Edinburgh, Royal Observatory, Blackford Hill, Edinburgh, EH9 3HJ U.K.}



\begin{abstract}

We present a study of the structural properties and metallicity distribution of the nearby peculiar galaxy NGC\,3077.  Using data from our survey of the M81 Group with the Hyper Suprime-Cam on the Subaru Telescope, we construct deep color-magnitude diagrams that are used to probe the old red giant branch population of NGC\,3077. We map these stars out to and beyond the nominal tidal radius, which allows us to derive the structural properties and stellar content of the peripheral regions.  We show that NGC\,3077 has an extended stellar halo and pronounced ``S-shaped" tidal tails that diverge from the radial profile of the inner region. The average metallicity of the old population in NGC\,3077 is estimated from individual RGBs to be $\gmh=-0.98 \pm 0.26$, which decreases with the distance from the galaxy center as $\gmh=-0.17$ dex $\rh^{-1}$. The metallicity of the S-shaped structure is similar to that of the regions lying at $r\sim4\times\rh (\sim 30$~kpc), indicating that the stellar constituents of the tidal tails have come from the outer envelope of NGC\,3077.  These results suggest that this peculiar galaxy was probably a rather normal dwarf elliptical galaxy before the tidal interaction with M81 and M82. We also examine the evidence in our dataset for the six recently-reported ultra-faint dwarf candidates around NGC\,3077. We recover a spatial overdensity of sources coinciding with only one of these. 

\end{abstract}


\keywords{galaxies: groups: individual (M81) --- galaxies: individual (NGC\,3077) --- galaxies: stellar content --- galaxies: structure}


\section{Introduction} 
NGC\,3077 is a member galaxy of the M81 Group and is located at a distance of 3.8 Mpc \citep{2001ApJ...555..280S}. This peculiar galaxy has been classified as an Irr II \citep{1961hag..book.....S}, Seyfert \citep{1968AJ.....73..866W}, I0pec \citep{1976srcb.book.....D}, and S0/Sa \citep{1981ApJ...243L..65B}.  \citet{1975gaun.book.....S} commented that NGC\,3077 is one of the few galaxies which cannot be accommodated by the Hubble scheme. The difficulty in classifying its morphology comes from the influence of the interaction with M81 and M82, which has distorted the original characteristics of NGC\,3077. \citet{1989ApJ...337..658P} concluded that NGC\,3077 was a small elliptical galaxy before the interaction, similar to the M31 dwarf satellite NGC\,185.

Together with M81 and M82, NGC\,3077 belongs to one of the most famous galaxy triplets. Conspicuous streams of atomic hydrogen gas (HI) result from tidal interactions in the past \citep{1994Natur.372..530Y,1999IAUS..186...81Y, 2018ApJ...865...26D} and both NGC\,3077 and M82 exhibit recent starburst activity in their central areas. It is rather surprising that NGC\,3077 has drawn much less attention from researchers in the past compared to M82. Based on a study of the resolved stellar populations of the M81 Group with Hyper Suprime-Cam (HSC) on the Subaru Telescope, \citet{2015ApJ...809L...1O} showed that the young stars closely follow the HI distribution and can be found in a stellar stream between M81 and NGC\,3077 and in numerous outlying stellar associations. 
In the central starburst region of NGC\,3077, young star clusters are also formed \citep{2004ApJ...603..503H}. X-ray observations with CHANDRA revealed that the central area of NGC\,3077 is dominated by a hot intercluster medium surrounded by H$\alpha$ super-shells, and the super-babbles have not yet exploded in contrast to the case of M82 \citep{2003ApJ...594..776O}.  Molecular gas clouds, as traced by CO \citep{2000A&A...361..500H,2001AJ....122.1770M,2002AJ....123..225W,2006AJ....132.2289W}, and the dust \citep{2011ApJ...726L..11W,2012A&A...543A..21H} are concentrated on the east side of NGC\,3077, coincident with the densest HI \citep{2006AJ....132.2289W}. Dust can also be seen in the main body of NGC\,3077, but the dust mass there is one order of magnitude smaller than that in the eastern area \citep{2011ApJ...726L..11W}. It is not clear whether the dense  HI, CO, and the dust were tidally stripped from NGC\,3077 or not \citep{2011ApJ...726L..11W}.

The HI tidal streams that link the three galaxies are believed to have formed during the last gravitational interaction between the three galaxies. The simulation by \citet{1999IAUS..186...81Y} showed that a close encounter between M81 and NGC\,3077 likely happened 280 Myrs ago, and another close encounter between M81 and M82 occurred  220 Myr ago. \citet{1999IAUS..186...81Y} failed to reproduce the stream between NGC\,3077 and M82 as the tidally disrupted outer disk of M81; instead, he was required to introduce a gas disk into the model for NGC\,3077. There is no observational evidence, however, that suggests the HI stream between M82 and NGC\,3077 comes from the tidally stripped gas from NGC\,3077. The central star-forming region suggests that the gas tidally stripped from M81 was accreted onto NGC\,3077 and induced star bursts \citep{1991MNRAS.252..543T}.  From nuclear emission line studies, \citet{1980A&A....87..142H} pointed out that the close companions of M81 (NGC\,3077, M82, and NGC\,2976) have abnormally high abundances of heavy elements for their absolute magnitude, suggesting that these galaxies might have borrowed metal-rich gas from M81 during close encounters. As \citet{1989ApJ...337..658P} and \citet{1991MNRAS.252..543T} pointed out, if NGC\,3077 was originally a dwarf spheroidal or elliptical galaxy, this would be incompatible with it having a disk of HI mass $\sim 10^9 \msun$ as \citet{1999IAUS..186...81Y} assumed in his simulation. 

From studying the color-magnitude diagrams (CMDs) and the luminosity functions, \citet{2019ApJ...884..128O} demonstrate that the star formation in outlying young stellar systems around M81 and NGC\,3077 has continued until very recently, roughly 30 Myr ago.  They also found that there are no over-densities of old red giant branch (RGB) stars associated with the star-forming region ``Garland'' and other young stellar systems, which led them to conclude that these systems are genuine young stellar systems whose star formation was triggered by the latest tidal interactions between M81, NGC\,3077, and M82. Although RGB stars at the location of Garland were identified in the Hubble Space Telescope (HST) imagery by \citet{2001ApJ...555..280S}, it is very likely that these RGB stars are halo stars of NGC\,3077.  Recently \citet{2022arXiv220909713Z} discovered the giant stellar stream associated with F8D1, an M81 satellite galaxy at roughly 2 degrees in projection from M81, which imply the tidal interactions of the M81 Group are more complicated than previously thought. 

The dynamical history of the M81-NGC\,3077 pair, and the M81 Group in general, remains unclear.  Numerical simulations of the M81-NGC\,3077 interaction by \citet{1993A&A...272..153T} showed that spiral arms and HI streams connecting M81 and NGC\,3077 are the results of the interaction between the two galaxies. \citet{2017MNRAS.467..273O} introduced dynamical friction in their M81 Group simulation, which suggests that three inner galaxies in the M81 Group are likely to merge within the next 1-2 Gyr. \citet{1999IAUS..186...81Y} also found in their simulation that three galaxies finally merged.  

\citet{2015ApJ...809L...1O} discovered a huge extended S-shape structure in the stellar component of NGC\,3077, which extends towards the north (the direction to M82) and the south.  Recently \citet{2022ApJ...937L...3B} identified five ultra faint dwarf (UFD) candidates of $\mv\sim-6$ around NGC\,3077, three of which seem to be embedded in this extended halo.  \citet{2020ApJ...905...60S} also showed that there is a significant amount of metal-rich stars currently unbound from NGC\,3077, and M82 due to the recent interaction, which may increase halo metallicity and mass of M81 in the near future.   

In this paper, we focus on the spatial structure and metallicity of the old resolved stellar population of NGC\,3077. Using data from our Subaru/HSC survey of the M81 Group, we map these properties across the main body of the galaxy as well as in its tidal extensions. Section 2 describes the data and CMDs. The structure and stellar populations are examined in Section 3 and Section 4, respectively. In Section 5, we discuss the metallicity, the origin of NGC\,3077, the survival timescale of S-shaped structure, and examine the UFD candidates of \citet{2022ApJ...937L...3B}. Finally, Section 6 gives a summary and the conclusions of our findings.

\section{Data and Color-Magnitude Diagrams} \label{sec: data}

\begin{figure*}
\begin{center}
 \includegraphics[width=500pt]{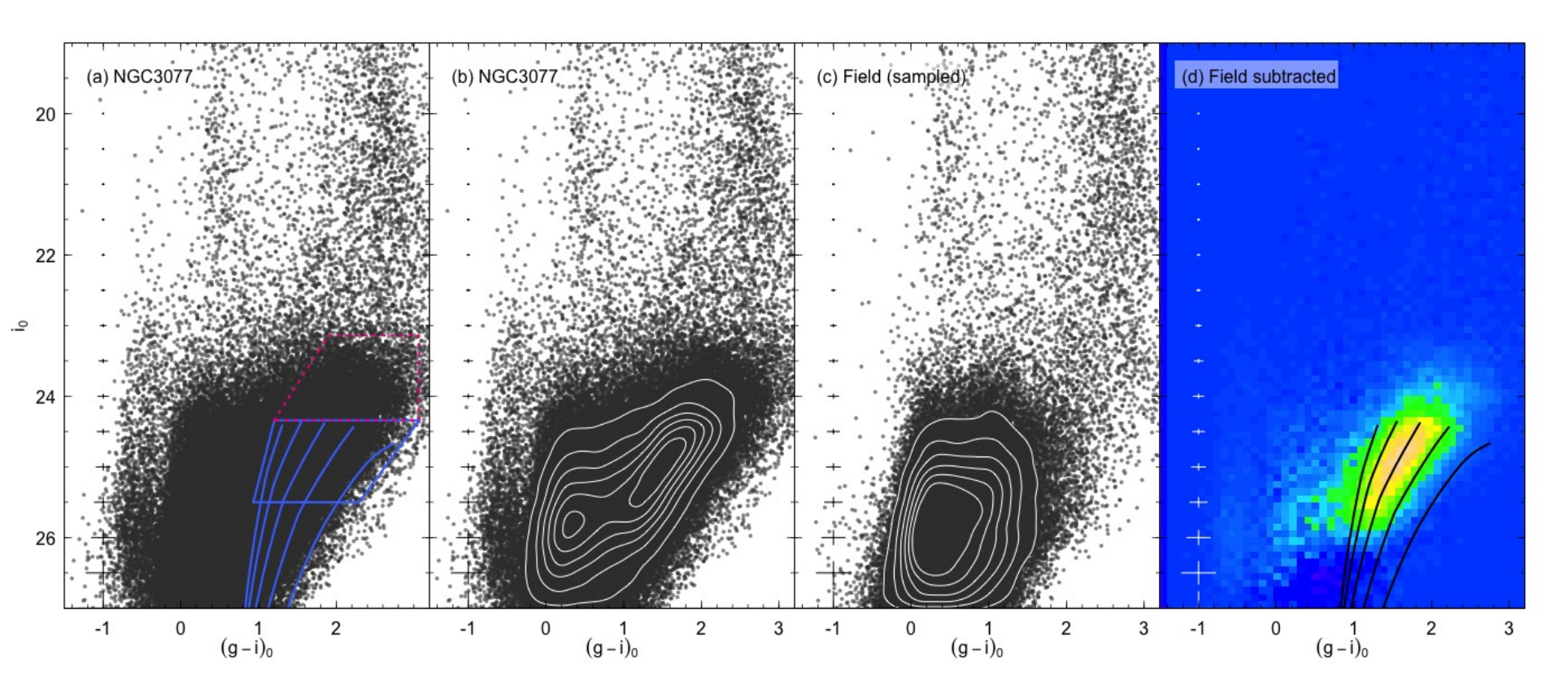}
 \vspace{10pt}
 \caption{De-reddened CMDs of stellar objects in NGC\,3077 and in the field.  (a) Stars within the tidal radius ($\rt$) of NGC\,3077.  Blue and magenta polygons indicate the selection criteria for RGB and AGB stars, respectively. PARSEC isochrones of age 10 Gyr and $\gmh=-2.2, -1.75, -1.3, -0.8, -0.5$ are overlaid as blue solid lines. The error bars show the mean photometric error of sources within $-1.5<\ogi<3.0$. (b) Same as CMD (a) but with density contours overlaid. {\bf Contours show number densities binned by 0.025 mag in the x-axis and 0.04 mag in the y-axis, and the levels are drawn every 50 from 50 to 300 stars per bin. } (c) Stellar objects in the field, selected as regions which lie outside of $4\times \rm{R}_{25}$ of M81 and M82, and outside of $\rt$ of the dwarf galaxies within the survey footprint.  The field counts are scaled so as to represent the same area as that contained within $\rt$ of NGC\,3077. The contour levels are the same with CMD (c).  (d) The field subtracted Hess diagram of NGC\,3077.  The binned difference between CMD (b) and CMD (c) is shown. PARSEC isochrones of age 10 Gyr are overlaid as black solid lines.} 
 \label{fig: cmds of field}
\end{center}
\end{figure*}

The data we analyze here are drawn from a photometric survey of the M81 Group that we have conducted with Hyper Suprime-Cam on the Subaru Telescope \citep{2012SPIE.8446E..0ZM}. Our survey covers 12 square degrees around M81 with seven pointings and reaches roughly two magnitudes below the RGB tip at the distance of M81.  The observing strategy, details of data processing and calibration have been previously described in \citet{2015ApJ...809L...1O, 2019ApJ...884..128O} and \citet{2022arXiv220909713Z}.  Throughout this study, we use the version of the de-reddened point-source catalog that is full-described in \citet{2019ApJ...884..128O}. To briefly recap, the HSC pipeline is used for the bias correction, flat-fielding, mosaicking, stacking, and calibration \citep{2018PASJ...70S...5B, 2019ApJ...873..111I, 2010SPIE.7740E..15A}.  For the dataset under study here, we then used the IRAF implementation of DAOPHOT \citep{1987PASP...99..191S} to obtain point spread function fitting photometry of sources. 
Artificial star tests were performed on some parts of the reduced images using the ADDSTAR task in DAOPHOT to add stars every 0.5 mag from mag=20.0 to 28.0.  The result is shown in Figure 2 of \citet{2019ApJ...884..128O}, which indicates the point source catalog is at least 50$\%$ complete to 26 mag in both passbands, except for the crowded region of NGC\,3077 center.  We fitted a modified sigmoid function to the point-source detection rate as functions of the magnitude and color, as given by the artificial star tests.  As seen by the hollow center in the left panel of Figure \ref{fig: rgb and contour map}, DAOPHOT  failed to detect sources in the highly crowded central few arc minutes of  NGC\,3077. To account for this, we only consider the r$>5\arcmin$ area for the analysis in this paper.
We used the DAOPHOT parameters $\chi^2$ and {\it sharpness} to separate point sources from the extended objects and noise-like detections.  Specifically, we select objects as stars whose $\chi^2$ and {\it sharpness} values lie within $3\times\sigma$ of the mean values of the artificial stars of the same magnitude.  

Figure \ref{fig: cmds of field} shows the CMDs of stellar objects in NGC\,3077 and in a reference field selected to be devoid of member stars of the M81 Group galaxies.  Black dots in panels (a) and (b) are de-reddened point sources that lie within the tidal radius of NGC\,3077 derived in Section \ref{sec: structure}.  In panel (a),  theoretical PARSEC v1.2S isochrones \citep{2012MNRAS.427..127B} of age 10 Gyr and metallicity $\gmh=-2.2, -1.75, -1.3, -0.8, -0.5$,  and of age 16, 32, 100 Myr with $\gmh=-0.75$, are overlaid as blue solid lines. In doing this, we have  assumed the distance modulus of $\mm=27.93$ to NGC\,3077 \citep{2001ApJ...555..280S}. The error bars show the mean photometric error of stars in $-1.5 < \ogi <  3.0$.  Panel (c) shows point sources that lie outside of $4\times \rm{R}_{25}$ of M81 and M82 \citep{1991rc3..book.....D}, and outside of $\rt$ of the dwarf galaxies that lie within the survey footprint studied in \citet{2019ApJ...884..128O}; these objects are expected to be contaminants, although they may contain a trace amount of stars in the extended halos of M81 and NGC\,3077. These objects are sampled so as to reflect the same area as that contained within $\rt$ of NGC\,3077.   Panel (d) is the Hess diagram showing the relative density of stars between panel (b) and panel (c). The bin size is 0.1mag.

A well-populated RGB is visualized by the extended contour at $\oi>24$ and at $\ogi>1$ in Figure \ref{fig: cmds of field}(b), which is not seen in Figure \ref{fig: cmds of field}(c).  The field-subtracted Hess diagram in Figure \ref{fig: cmds of field}(d) also shows this overdensity.  Another striking feature in Figure \ref{fig: cmds of field}(b) is the young blue population vertically distributed at $\ogi<-0.4$.  We select the bright RGB population using the solid blue polygon shown in Figure \ref{fig: cmds of field}(a). The boundaries of this have been chosen to optimize the selection of genuine NGC\,3077 stars while limiting the number of foreground and background contaminants.  We note that there is a substantial asymptotic giant branch (AGB) population within the dotted magenta area above the RGB selection box. As discussed in \citet{2015ApJ...809L...1O}, these bright AGB stars are centrally concentrated in NGC\,3077, which implies these stars are intrinsic populations.  Further investigation on the AGB population of NGC\,3077 will be done in the future with high-resolution NIR imagery from Euclid.

\section{The Outer Structure of NGC\,3077} \label{sec: structure}

\begin{figure*}
\begin{center}
 \includegraphics[width=500pt]{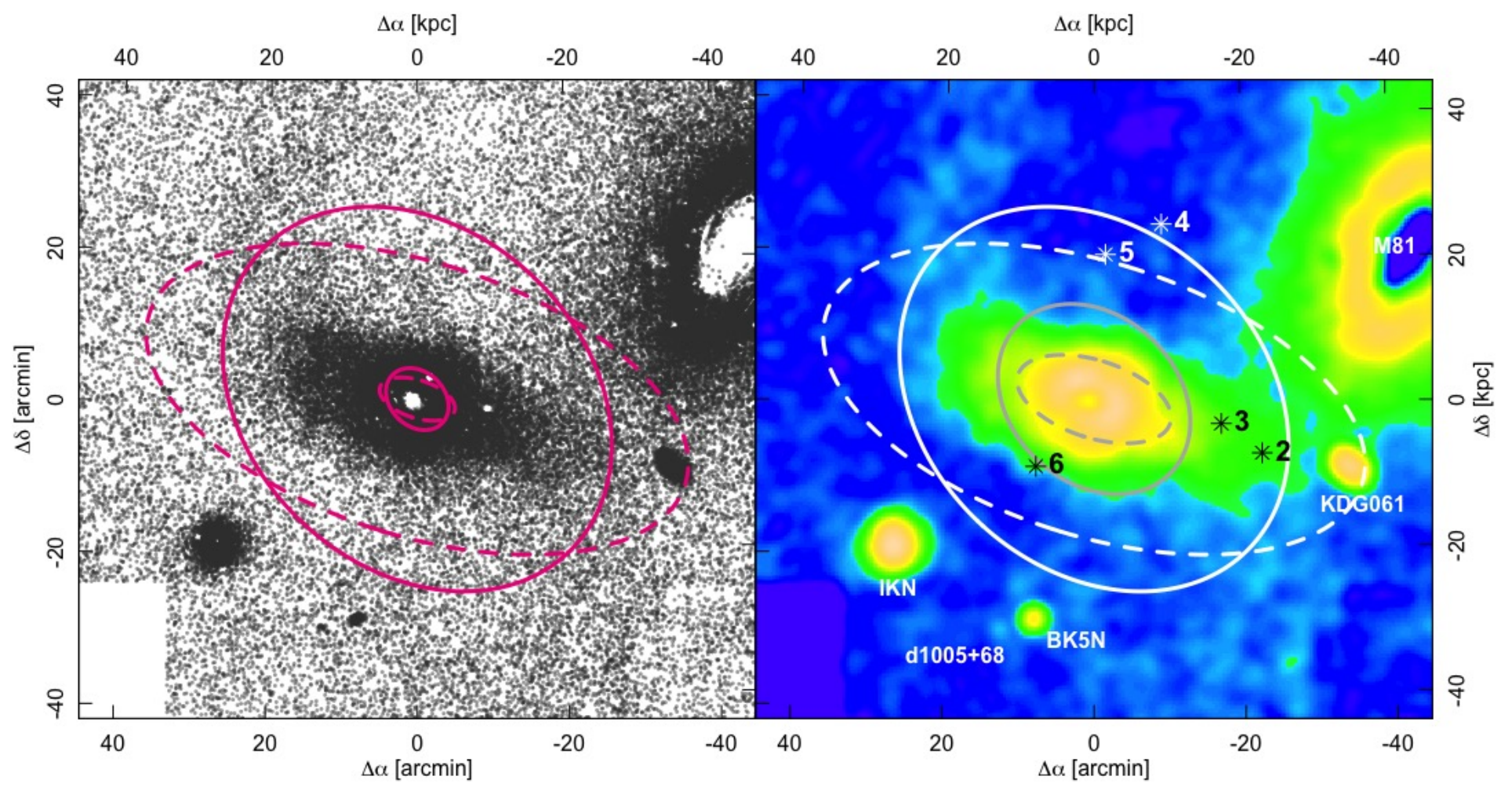}
 \vspace{10pt}
 \caption{{\it Left:} The spatial distribution of RGB stars in and around NGC\,3077. The solid ellipses represent the King profile parameters $\rc$ and $\rt$ as estimated using star counts over the whole area, while the dashed lines show those parameters derived considering only stars in the outer region.  {\it Right:} A smoothed density map of RGB stars shown on a logarithmic scale and reflecting densities starting at  $0.4\sigma$ of the background level.  The kernel density is estimated with the bandwidth of $1\arcmin.0$. Other M81 Group galaxies projected nearby are labeled with white letters. The white solid and dashed lines represent $\rt$ as the same as the left panel.  The gray solid and dashed lines show the break radii $\rb=14\arcmin$ and $\rm{R_{B,E}}=10.5\arcmin$ (see Section \ref{subsec: timescale of S-shape}). White and black star symbols indicate the locations of UFD candidates reported by \citet{2022ApJ...937L...3B} (see Section \ref{subsec:ufds}). }
 \label{fig: rgb and contour map}
\end{center}
\end{figure*}
Using the RGB stars selected from the CMD, we proceed to analyze the structure of NGC\,3077 in the same manner as done in \citet{2019ApJ...884..128O}. Briefly, this involves deriving the centroid, position angle, and ellipticity using the density-weighted first and second moments of the RGB star spatial distribution and then using these parameters to construct the radial star count profile.  

As seen in Figure \ref{fig: rgb and contour map}, NGC\,3077 has an extended and diffuse stellar halo and pronounced ``S-shaped" tidal tails. The halo exhibits a rather boxy shape, which is not well-matched by the position angle and the ellipticity derived from the fit to the full star count distribution (solid ellipses).   We also derive the average ellipticity and position angle of the star counts excluding the inner r$=8.4\arcmin$ area of NGC\,3077. The parameters derived in this way provide a much better description of the outer structure of NGC\,3077 as shown by the dashed lines in Figure \ref{fig: rgb and contour map}. The resulting structural properties are listed in Table \ref{tbl: str}. Hereafter we use $\rm{R}_{{\rm{(out)}}}$ to denote radii estimated using the ellipse parameters derived from only the stars in the outer region. 

\begin{figure}
\begin{center}
 \includegraphics[width=240pt]{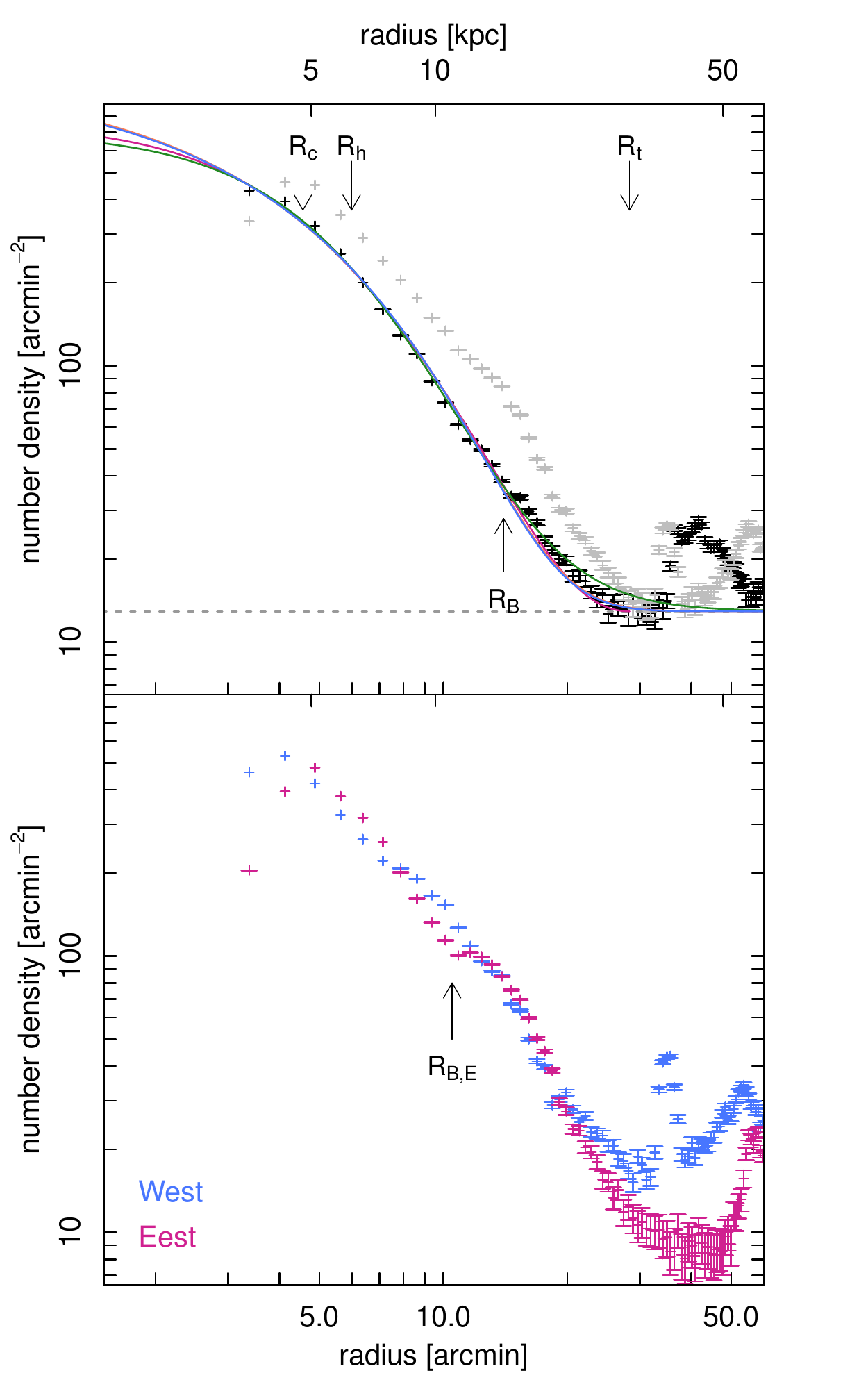}
 \caption{Completeness-corrected radial number density profile of RGB stars in and around NGC\,3077. {\it Upper panel}: Azimuthally-averaged radial star count profiles derived within elliptical annuli.  The best-fitting King, exponential, Plummer, and Sersic profiles are overlaid as magenta, orange, green, and blue lines, respectively, although the exponential profile (orange line) is almost overlapped with the Sersic profile (blue). The foreground/background contamination level is shown as a dashed horizontal line.  The profile derived using the ellipse parameters derived from stars in the outer region only is shown as a gray color.  {\it Lower panel}: Radial profiles for the west and east sides of NGC\,3077 using structural parameters derived from stars in the outer region are shown as blue and magenta colors, respectively. }
 \label{fig: radial profile}
\end{center}
\end{figure}

\begin{deluxetable*}{rcccccccccc}
\tabletypesize{\scriptsize}
\tablecolumns{10}
\tablewidth{0pt}
\tablecaption{The structural properties of NGC\,3077 \label{tbl: str}}
\tablehead{
\colhead{Galaxy}  & \colhead{$\alpha$} & \colhead{$\delta$}   & \colhead{P.A.$^{\rm{(a)}}$}  & \colhead{$\epsilon$$^{\rm{(b)}}$}   &
\colhead{$\rc$$^{\rm{(c)}}$} & \colhead{$\rt$$^{\rm{(d)}}$} & \colhead{$\re$$^{\rm{(e)}}$} & \colhead{b$^{\rm{(f)}}$} & \colhead{n$^{\rm{(g)}}$} & \colhead{$\rh$$^{\rm{(h)}}$}\\
 & (J2000) & (J2000) & (deg) & & (arcmin) & (arcmin) & (arcmin) & (arcmin) & & (arcmin)\\
 & & & & & (kpc) & (kpc) & (kpc) & (kpc) & & (kpc)}
\startdata
NGC\,3077 & $10:03:14.2$ & $+68:44:19.9$ &  $46.3$ & $0.22$ & $4.6\pm0.2$ & $28\pm1$ & $3.6\pm0.1$ & $6.7\pm0.1$ & $1.0\pm0.1$ & $6.0\pm0.1$ \\
 & & & & & $5.1\pm0.2$ & $32\pm2$ & $4.0\pm0.1$ & $7.5\pm0.1$ & & $6.7\pm0.1$ \\
NGC\,3077 (outer) &      &               &  $72.3$ & $0.52$ & $5.1\pm0.2$ & $37.0\pm1.2$ & $4.31\pm0.07$ & $8.2\pm0.2$   & $1.18\pm0.06$ & $7.2\pm0.1$ \\
 & & & & & $5.7\pm0.2$ & $41\pm1.3$ & $4.8\pm0.1$ & $9.1\pm0.2$ & & $8.1\pm0.1$\\
\enddata
\tablecomments{(a) Position angle from north to east. (b) Ellipticity $\epsilon=1-\rm{b/a}$ where a is the scale length of the galaxy along the major axis and b is that along the minor axis. (c) Core radius of King profile. (d) Tidal radius of King profile. (e) Scalelength of the exponential profile. (f) Plummer radius. (g) Sersic index. (h) Half light radius.}
\end{deluxetable*}

The completeness-corrected radial number density distribution of RGB stars is fits with the standard King, exponential, Plummer, and Sersic models via least-squares minimization. The best-fit profiles are overlaid as magenta, orange, green, and blue lines, respectively, in the upper panel of Figure \ref{fig: radial profile}. The foreground/background contamination level is shown as a dotted horizontal line. Our imagery does not resolve stars in the very inner part of NGC\,3077, as evidenced by the hole in the left panel of Figure \ref{fig: rgb and contour map}, hence we exclude the inner $5\arcmin$ region from the fit.  The rise in the profile beyond $\rt$ is due to contamination from the stellar components of M81 and the other dwarf galaxies in the surrounding area.  The star count profile is generally well-described by all the models.  That said, there is evidence for a clear kink in the radial profile at around $\rm{r}\approx14\arcmin$.  The radial profile derived using the structural parameters of the outer region is shown in the gray color in the upper panel of Figure \ref{fig: radial profile}, where the kink is even more obvious. This radius is indicated with the gray solid ellipse in the right panel of Figure \ref{fig: rgb and contour map}. Such features have been suggested to reflect the break radius $\rb$ at which a transient excess in the stellar density profile appears due to tidal stripping of a dwarf galaxy \citep{2002AJ....124..127J, 2009ApJ...698..222P}. The stars in these features have gained energy at the last pericentric passage and are moving outwards as they settle into new, less bound orbits in the dwarf.  We will use this $\rb$ to estimate the elapsed time since NGC\,3077 last pericentric passage around M81 in Section \ref{subsec: timescale of S-shape}.    

\begin{figure}
\begin{center}
 \includegraphics[width=240pt]{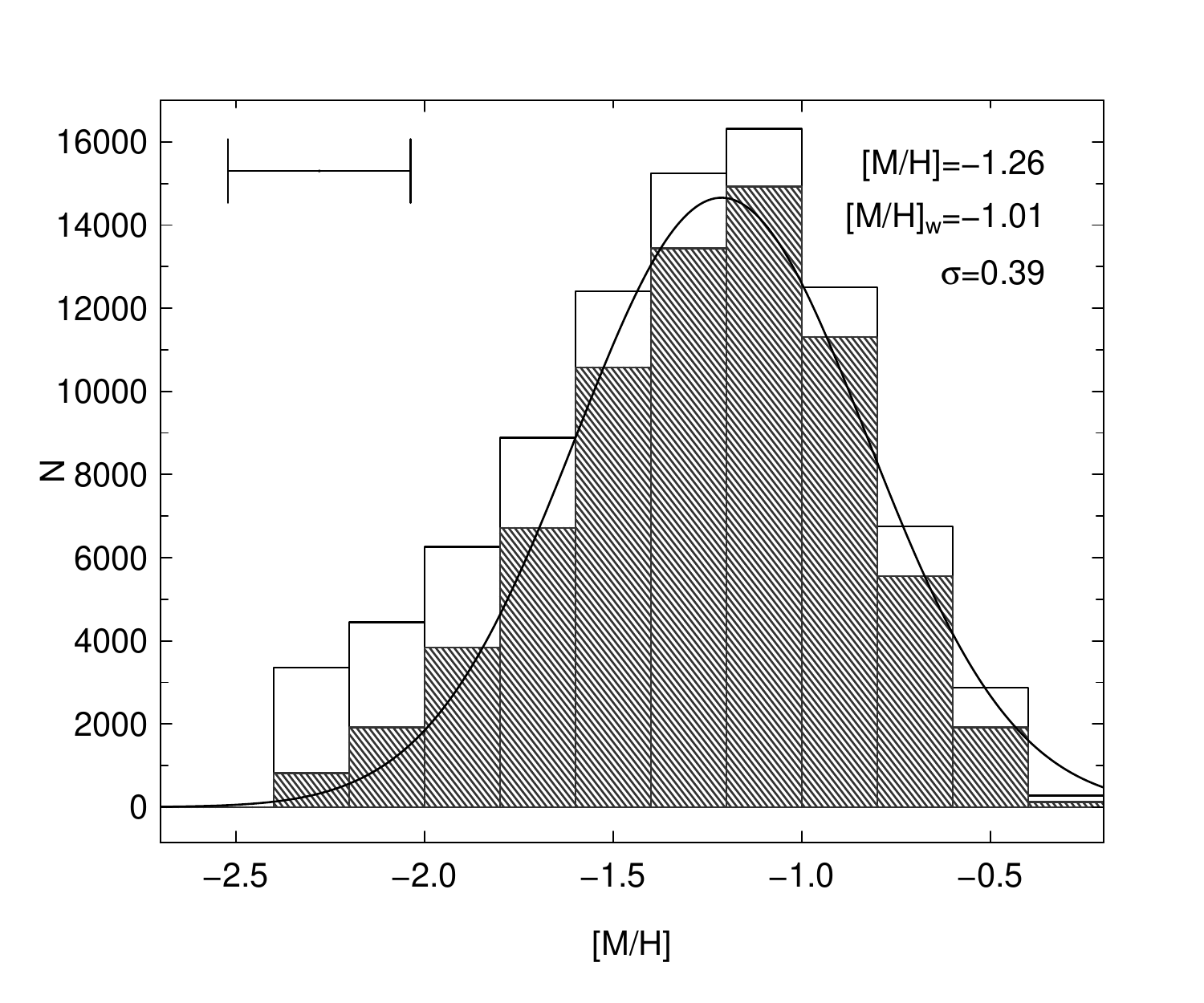}
 \vspace{10pt}
 \caption{Metallicity distribution of RGB stars within $4\times\rh_{\rm{(out)}}$ of NGC\,3077 derived using the $\ogi$ colour and assuming a 10 Gyr population. {\bf The black open histogram shows the completeness-corrected MDF using all RGB stars. The shaded histogram is the field-subtracted MDF. The solid line shows the fitted single Gaussian distribution to the field-subtracted MDF. The mean metallicity error is shown on the upper left side.}}
 \label{fig: mdf}
\end{center}
\end{figure}

To further investigate the bump feature, we divide the area around NGC\,3077 into eastern  and western halves.  The lower panel of Figure \ref{fig: radial profile} shows the radial stellar density of the eastern and the western sides. The radial star counts of each  are constructed by calculating the completeness-corrected average number of RGB stars with $\Delta\rm{RA}>0$ ($\Delta\rm{RA}<0$) in elliptical annuli, defined using the structural parameters derived from stars in the outer region. The profile of the western  side is highly contaminated by the presence of M81, while the distribution in the eastern side shows a pronounced bump at $\rm{R_{B,E}} \approx 10.5 \arcmin$. The gray dashed ellipse in Figure \ref{fig: rgb and contour map} shows that this also corresponds to the location where the structure of the dwarf starts to become increasingly elongated.

\section{The Metallicity Distribution of the Old Population} \label{sec: pop}

\begin{figure}
\begin{center}
 \includegraphics[width=240pt]{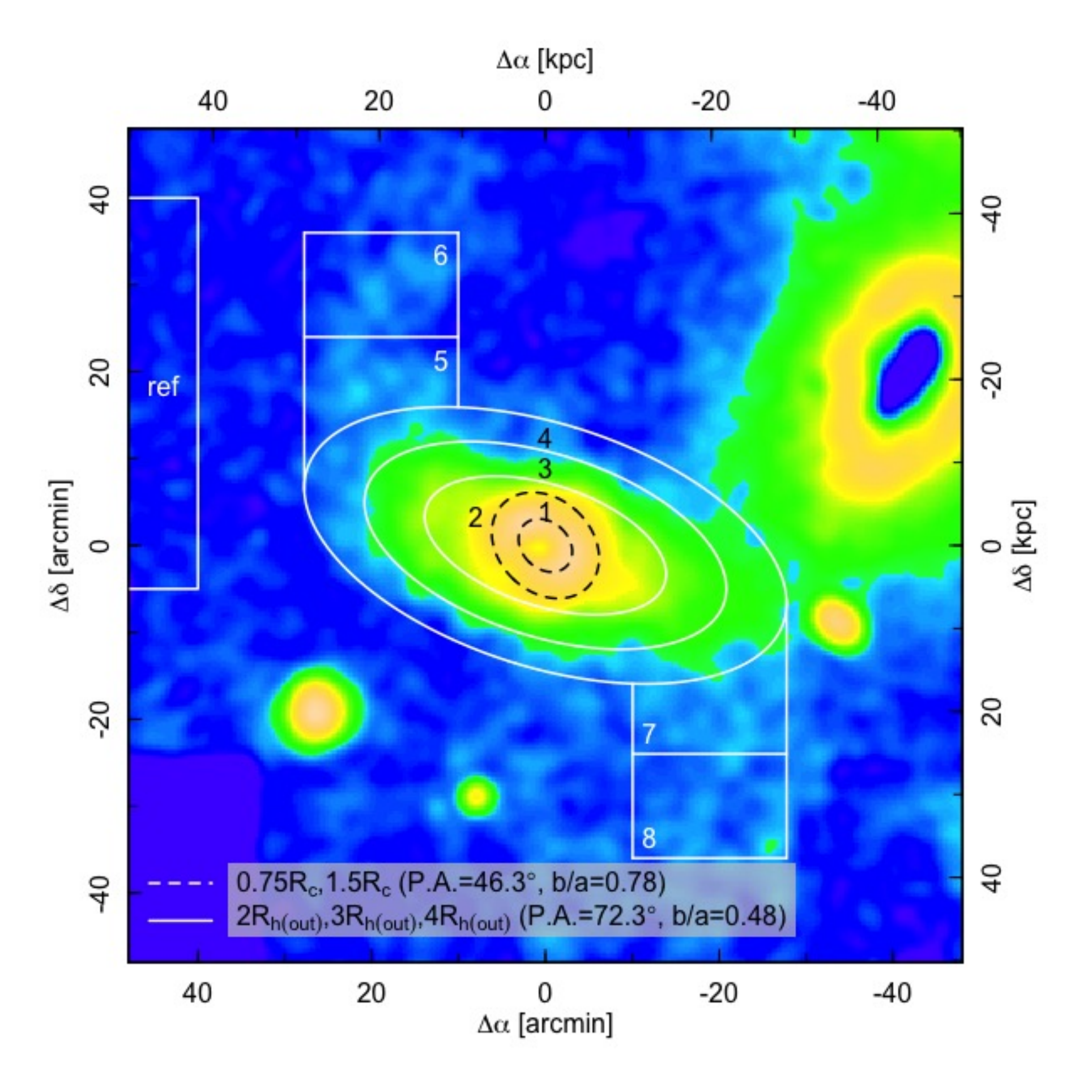}
 \caption{The smoothed density map of RGB stars shown on a logarithmic scale on which is overlaid dashed ellipses indicating  $0.75\times\rc$ and $1.5\times\rc$ radii and 
 solid ellipses indicating $2\times\rh_{\rm{(out)}}, 3\times\rh_{\rm{(out)}},4\times\rh_{\rm{(out)}}$.  CMDs of stars located within the nine labeled regions are shown in Figure \ref{fig: cmd9}. }
 \label{fig: map9}
\end{center}
\end{figure}

\begin{figure*}
\begin{center}
 \includegraphics[width=410pt]{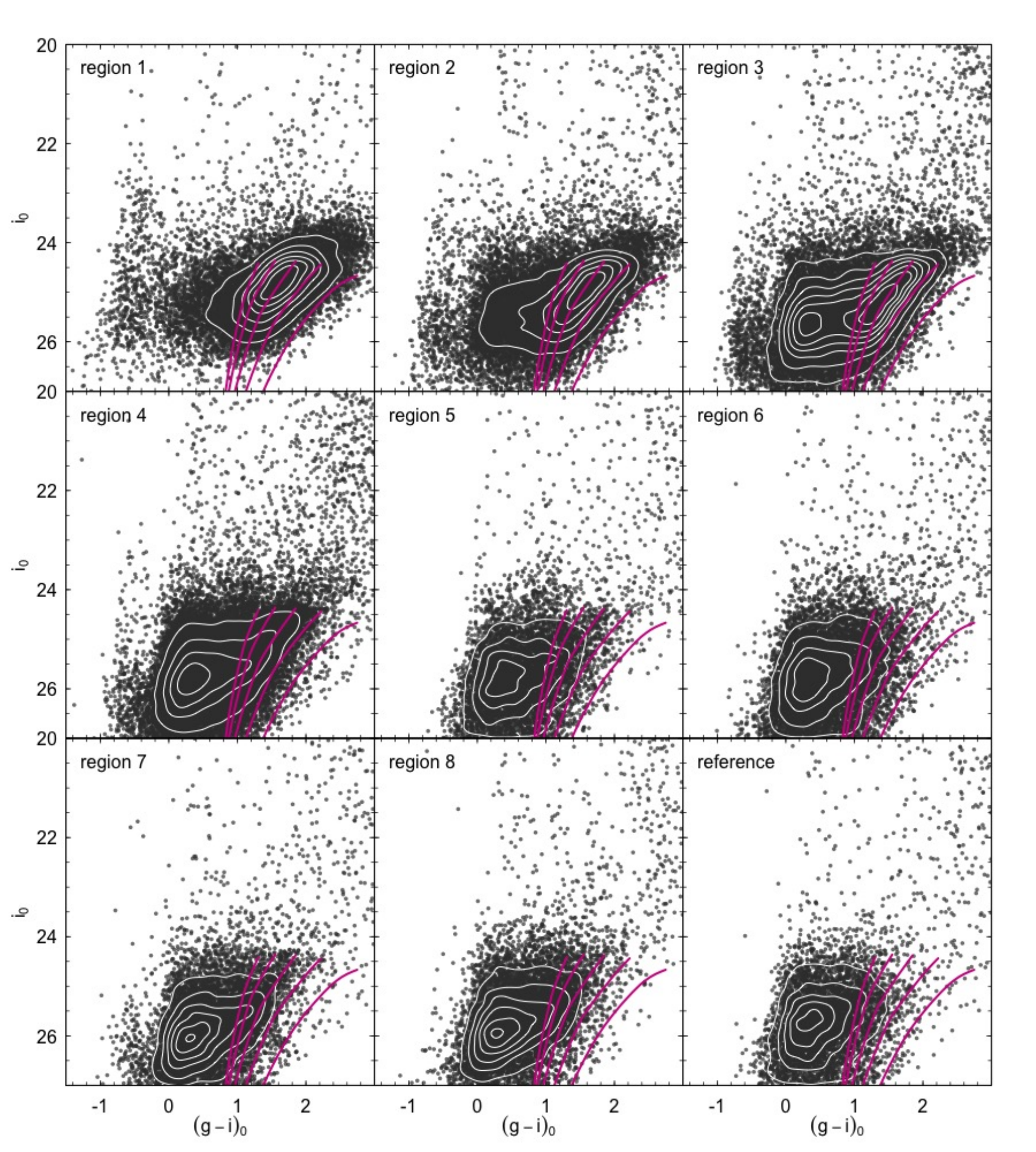}
 \vspace{10pt}
 \caption{De-reddened CMDs of stellar objects within nine regions defined in Figure \ref{fig: map9}.  The CMD of the reference region is scaled to reflect the same areal coverage as that of region 6 and region 8.  The Parsec v1.2S isochrones of 10 Gyr with $\gmh=-2.2, -1.75, -1.3, -0.8, -0.5$ are overlaid as magenta solid lines. }
 \label{fig: cmd9}
\end{center}
\end{figure*}

In this section, we examine the metallicity distribution of RGB stars in the inner and outer regions of NGC\,3077, which we have established exhibit different structural properties in the previous section.  The black open histogram in Figure \ref{fig: mdf} is the metallicity distribution function (MDF) of all RGB stars lying within $4\times\rh_{\rm{(out)}}$ of the NGC\,3077 center, corresponding to the outermost ellipse shown in Figure \ref{fig: map9}.  We derive the metallicity of individual RGB stars using the $\ogi$ color and assuming a 10 Gyr population in the same manner as those in \citet{2019ApJ...884..128O}.

 The photometric completeness of each RGB star is corrected according to its color and magnitude. Foreground and background contamination is estimated using point sources drawn from the reference region shown in Figure \ref{fig: map9}, which is scaled so as to represent the same area. The field contamination corrected MDF is shown as the shaded histogram in Figure \ref{fig: mdf}.  The metallicity errors of individual RGB stars are estimated by performing a Monte Carlo simulation with N$=2000$ for each star, sampling from a Gaussian distribution with a width equal to the photometric error.  The estimated uncertainties increase from $0.09$ to $0.48$ with the decreasing metallicity from $\gmh=-0.3$ to $-2.5$ due to the narrow range of RGB color between metal-poor isochrones.  The mean metallicity corrected for this effect is shown as the error-weighted mean value $\gmh_{\rm{W}}=-0.98 \pm 0.24$. The error bar in Figure \ref{fig: mdf} indicates the uncertainty at the mean metallicity. 
 Our analysis uses all stars within the RGB box but experiments showed that the results did not change if we instead only used the brightest half of the RGB.

\begin{figure*}
\begin{center}
 \includegraphics[width=500pt]{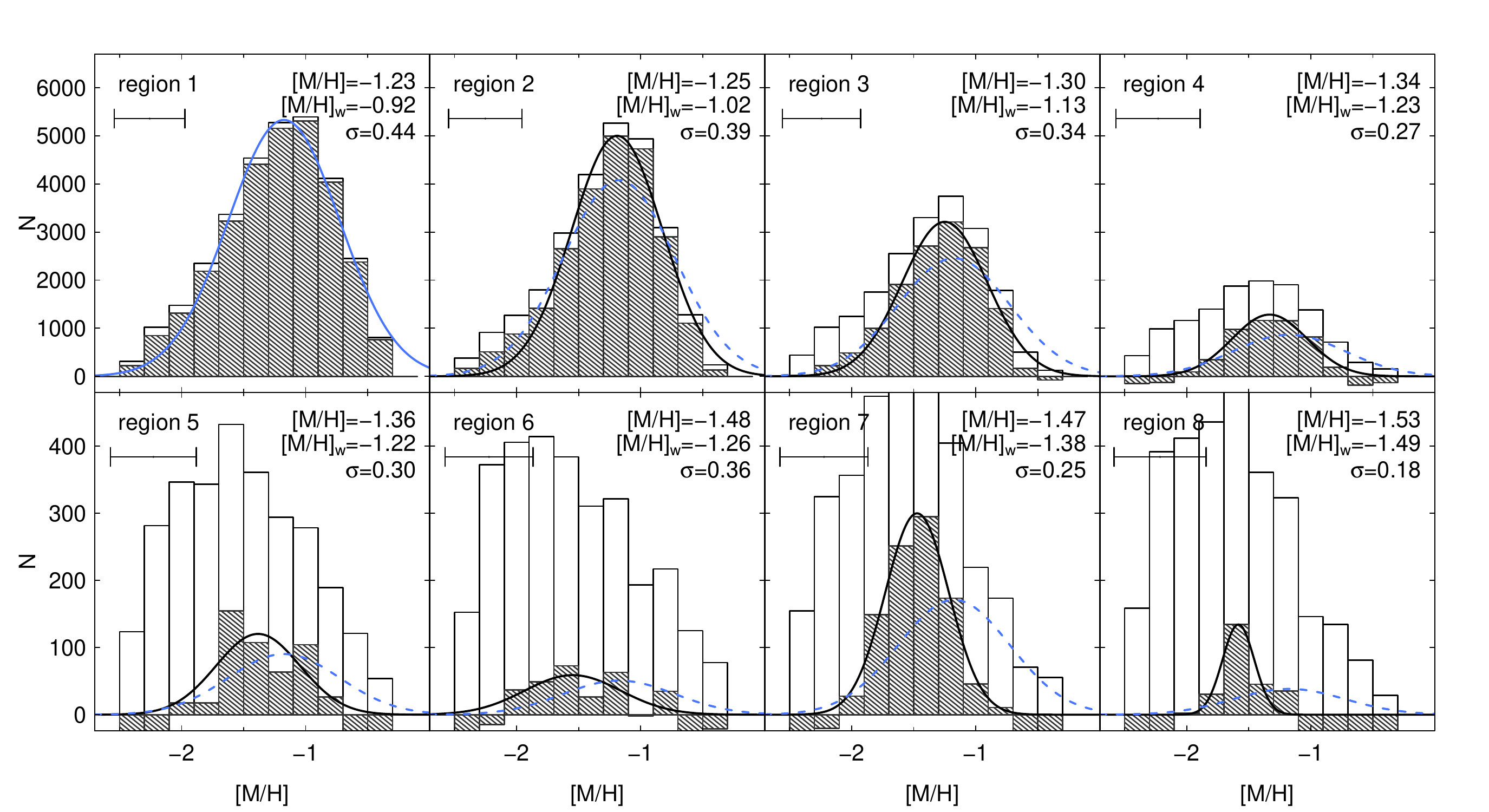}
 \vspace{10pt}
 \caption{MDFs of RGB stars in regions 1 to region 8, as defined in Figure \ref{fig: map9}. The shaded histograms are the field-subtracted MDFs. The solid lines show single Gaussian fits to the distributions.  The blue dashed lines in regions 2-8 are the scaled versions of the Gaussian fit to region 1 for reference purposes. The region ID, mean metallicity, weighted mean metallicity, and dispersion are listed in each panel. }
 \label{fig: mdf8}
\end{center}
\end{figure*}
 
To compare the MDFs in different areas in the galaxy, we divide NGC\,3077 into eight regions; (1) $0.75~\rc<r<1.5~\rc$, (2) $1.5~\rc<r<2~\roh$, (3) $2~\roh<r<3~\roh$, (4) $3~\roh<r<4~\roh$, (5) and (6) north part of the S-shaped structure, (7) and (8) south part  of the S-shaped as shown in Figure \ref{fig: map9}.  The CMDs of stars within each region are shown in Figure \ref{fig: cmd9}.  In regions 1 to 3, the dominant populations are seen to be RGB stars with $1.0<\ogi<2.0$ and $\oi>24.0$.  This RGB population diminishes in prominence moving from the inner to outer regions, and unresolved background galaxies with $\ogi\sim0.4$ and $\oi\sim25.8$ are the dominant feature in  the outer ``S-shaped" structure  CMDs (regions 5-8) and the reference CMD. Vertical features corresponding to young populations lie at $\ogi<-0.4$ and are visible in regions 1 to 4. The only evidence of young stars in the ``S-shaped" feature comes from region 7, where the young stellar stream linking NGC\,3077 and M81 was discovered \citep{2015ApJ...809L...1O}.  Figure \ref{fig: cmd9} also shows that the luminous intermediate-age AGB population, which is located above the RGB tip, is also less prominent, moving from the center to the outer regions. 

Figure \ref{fig: mdf8} shows the MDF of the RGB populations in each region, derived in the same manner as for Figure \ref{fig: mdf}. The foreground and background counts in the reference field are scaled so as to represent the same area as each region.  The blue dashed lines in the panels of regions 2-8 are the scaled versions of the Gaussian fit to region 1 and are shown for reference.  Moving from the inner to the outermost parts of the main body (regions 1 to 4), the mean metallicity and metallicity dispersion decrease with radius.  Though it is within the error bars, a mild radial metallicity gradient can be derived as $\Delta\gmh=-0.02$ dex/kpc in physical scale or $-0.14$~dex $\roh^{-1}$ as a function of the half-light radius using the structural parameters derived for the outer region.  In  the ``S-shaped" feature (region 5-7), the contaminants dominate over the RGB population, but the reference field-subtracted histograms still show a comparable or slightly metal-poor population with that of region 4. This is consistent with this structure being made up of stars that have been stripped from the outermost part of NGC\,3077 due to the tidal interaction with M81 and M82. 


\section{Discussion} \label{sec: discussion}

\subsection{Metallicity}
\label{subsec: gradient}

In the previous section, we derive a mild metallicity gradient in the old RGB population of NGC\,3077 in the sense of decreasing metallicity with increasing radius.  Such gradients are common in Local Group dwarf galaxies \citep{2022A&A...665A..92T}. NGC\,3077's gradient, $-0.14$ dex $\roh^{-1}$ is comparable with that of the dwarf elliptical NGC\,185, $-0.2$ dex $\rh^{-1}$ \citep{2022A&A...665A..92T}.  The MDF is also similar to that of NGC\,185 \citep{2014MNRAS.445.3862C}. Although NGC\,3077 has much larger radius of $\roh=8.1$ kpc than NGC\,185 (0.5~kpc) due to the recent gravitational interaction with M81 and M82, these results support the suggestion by \citet{1989ApJ...337..658P} that NGC\,3077 was once a normal dwarf elliptical, akin to  NGC\,185 before its interaction.

The presence of metallicity gradients in dwarf galaxies has been examined in high-resolution cosmological simulations of galaxy formation. Recently \citet{2021MNRAS.501.5121M} use the FIRE-2 simulation suite to investigate the presence of metallicity gradients in dwarf galaxies of stellar mass $10^{5.5}$ to $10^{8.8}~\msun$. They show that inverse metallicity gradients in the stellar component are common and are set by the competition between the ``puffing'' process of old, metal-poor stars by feedback-driven potential fluctuations in the early phases of formation and the late time accretion of metal-rich gas that provides fuel for extended star formation.  In this sense, NGC\,3077's  metallicity gradient may eventually be washed out due to new-born metal-rich stars formed in the peripheral regions by gas displaced  due to the recent tidal interactions. 

\begin{figure}
\begin{center}
\includegraphics[width=240pt]{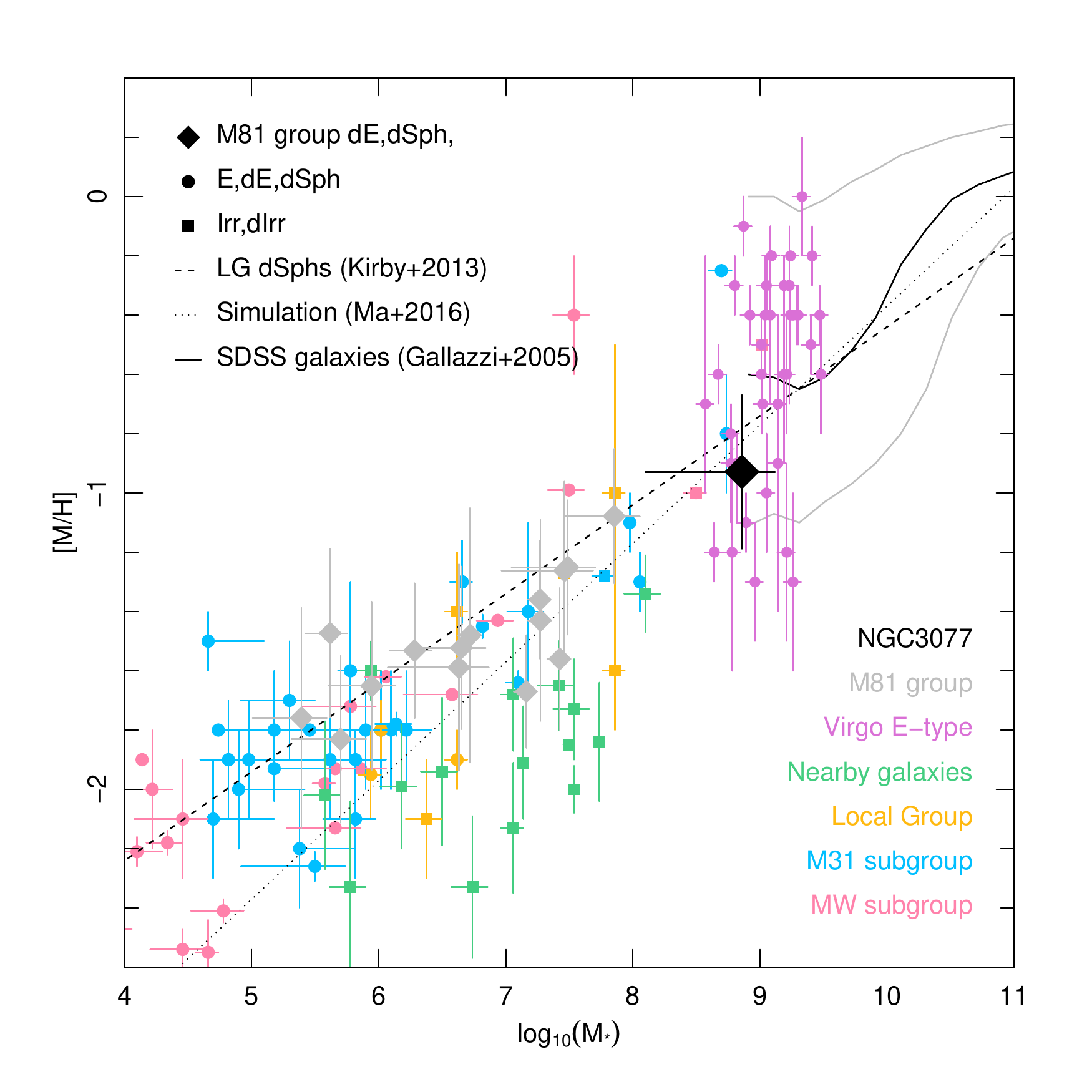}
\caption{The stellar mass-metallicity relation of nearby galaxies. NGC\,3077 is shown with a black diamond. The other galaxies shown include the M81 dwarf galaxies studied in \citet{2019ApJ...884..128O} and \citet{2010A&A...521A..43L}, nearby dwarf  galaxies from \citet{2012AJ....144....4M}  and early-type dwarf galaxies in the Virgo cluster from \citet{2014ApJS..215...17T}.  Where necessary, we have assumed the M/L of \citet{2008MNRAS.390.1453W} to derive stellar masses.}
\label{fig: dwarfs Mv metal Rh}
\end{center}
\end{figure}

Figure \ref{fig: dwarfs Mv metal Rh} shows the stellar mass and metallicity of NGC\,3077 compared to that of other galaxies in the local universe. The M81 dwarf galaxies studied in \citet{2019ApJ...884..128O} and \citet{2010A&A...521A..43L}, nearby dwarf galaxies from \citet{2012AJ....144....4M} and early-type dwarf galaxies in the Virgo cluster from \citet{2014ApJS..215...17T} are plotted.  We convert galaxies $\mv$ in the literature to the stellar mass, assuming the M/L of dE and dIrr in \citet{2008MNRAS.390.1453W}.  The mass-metallicity relations of the Local Group galaxies \citep{2013ApJ...779..102K}, FIRE simulation suite \citep{2016MNRAS.456.2140M}, and SDSS galaxies \citep{2005MNRAS.362...41G} are also plotted in Figure \ref{fig: dwarfs Mv metal Rh}. NGC\,3077's stellar mass of $7.2\pm1.2 \times10^{8} \msun$ is calculated using RGB and AGB stars within $4\times\roh$ (the outermost ellipse in Figure \ref{fig: map9}) to avoid contamination of M81 and nearby satellites.  The $i$-band and $g$-band luminosities of individual RGB and AGB stars located between $0.75\times\roc$ to $4\times\roh$ are summed up with the photometric completeness correction. The total luminosity is corrected to account for the flux of inner $0.75\times\roc$ area using the extrapolated Sersic radial profile and for the flux of stars fainter than the RGB selection box using the 10 Gyr and $\gmh=-1.3$ luminosity function, then it is corrected the background/foreground contamination using the reference field. The total magnitude $\mv$ is calculated from transforming $\mg$ and $\mi$ \citep{2006A&A...460..339J}, and then converted to the stellar mass assuming $\mv/L=1.6$ \citep{2008MNRAS.390.1453W}. Considering just the old stellar components of NGC\,3077, as we have done here, it can be seen that NGC\,3077 lies very close to the mass-metallicity relation defined by dwarf galaxies and is very close to dwarf ellipticals in the Virgo cluster.   

\subsection{Origin of NGC\,3077}

Various pieces of evidence support the idea that NGC\,3077 was an NGC\,185-like dwarf elliptical galaxy before the interaction with M81 and M82.  This includes the fact that its radial stellar density profile is well-fit by a King and other models which are traditionally used to describe dE/dSph. It is also located on the mass-metallicity relation defined by nearby galaxies and inhabits a region of the diagram that is populated by early-type dwarfs. Finally, it has a stellar MDF that is very similar to those of low-mass elliptical galaxies \citep{2014MNRAS.445.3862C}. 
Taken together, it seems unlikely that NGC\,3077 was originally a late-type dwarf with an extended gas disk as \citet{1999IAUS..186...81Y} assumed in his numerical simulation.  However, without this assumption,  \citet{1999IAUS..186...81Y} could not reproduce the global distribution of HI  in the M81 Group. This implies that there may yet be missing ingredients in the modeling of the interaction history of the M81 Group. The recent discovery of a giant tidal stream from the dwarf galaxy F8D1 \citep{2022arXiv220909713Z} suggests that it also had a recent passage close to M81 and may also have been stripped of its gas at this time. Further modeling of how interactions can explain the extended HI gas in the M81 Group is clearly warranted. 

\begin{figure}
\begin{center}
\includegraphics[width=240pt]{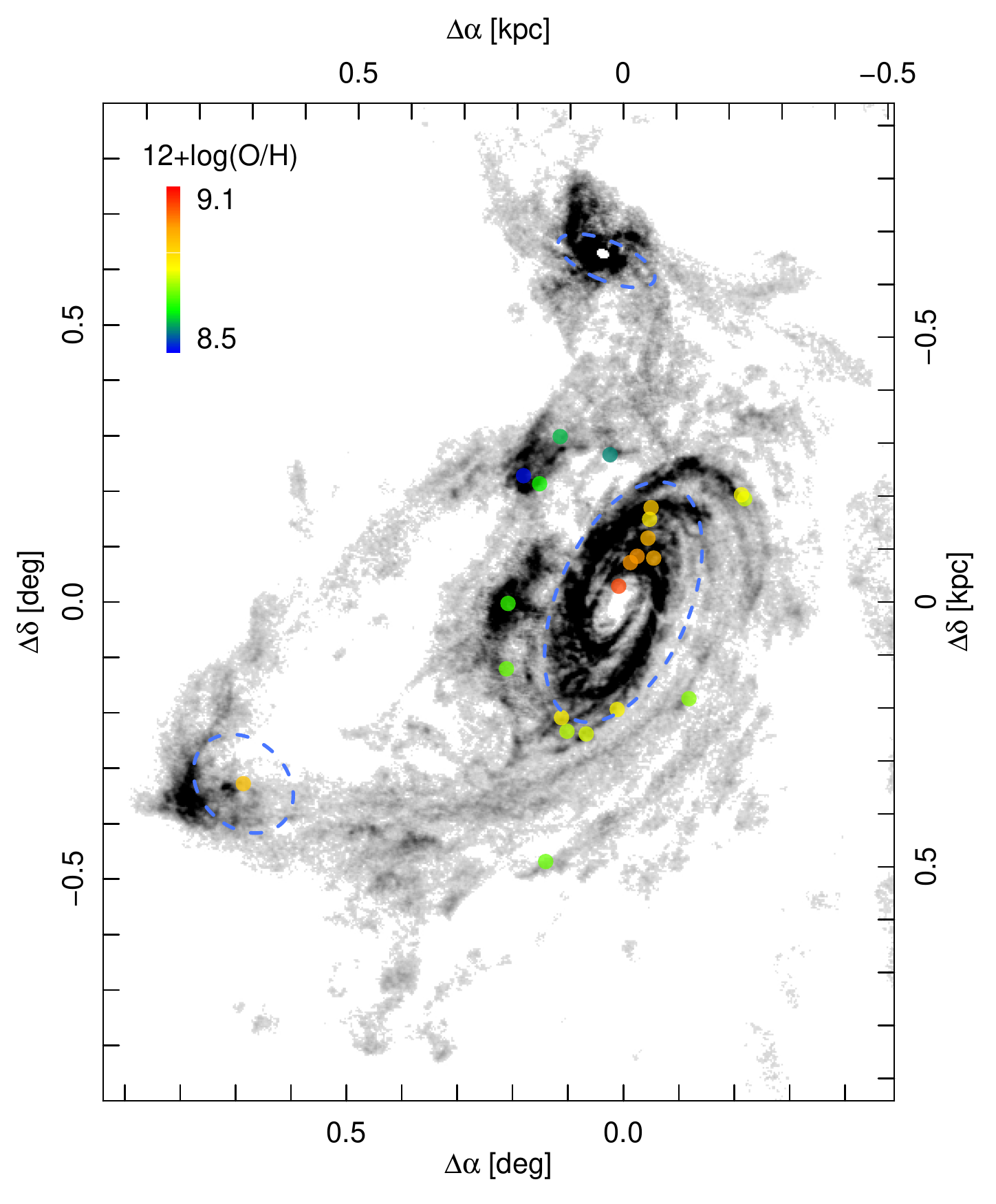}
\caption{The oxygen abundance of M81 and NGC\,3077 on the HI map taken from \citet{2018ApJ...865...26D}.  The metallicity of 21 regions around M81 from \citet{2012MNRAS.422..401P} and at the center of NGC\,3077 from \citet{2004AJ....127.1405C} are color-coded according to the metallicity.}
\label{fig: gas metal}
\end{center}
\end{figure}

Another argument against the gas around NGC\,3077 being of intrinsic origin comes from consideration of the gas-phase metallicity. The gas oxygen abundance of NGC\,3077 is much higher than expected from its luminosity \citep{1980A&A....87..142H}. 
 Figure \ref{fig: gas metal} shows the oxygen abundances that have been measured for M81 and NGC\,3077 superposed on the HI map of \citet{2018ApJ...865...26D}.  \citet{2004AJ....127.1405C} derived $\ohmetal=8.9\pm0.1$ for NGC\,3077 by applying the \citet{2002ApJS..142...35K} calibration to measurements of the strong emission lines. By using the same \citet{2002ApJS..142...35K} calibration, \citet{2012MNRAS.422..401P} derived oxygen abundances in the gas disk of M81 and found a radial gradient going from $\ohmetal\sim9.0$ at the galaxy center to $\sim8.6$ at 30 kpc. This strongly suggests that the gas currently in the inner regions of NGC\,3077 was once part of the outer disk of M81 and accreted onto NGC\,3077 as a result of the tidal interaction. In this scenario, the HI stream that links M81 and NGC\,3077 represents the extension of a tidal arm that resulted from NGC\,3077 falling in on a prograde orbit within the plane of M81's disk \citep{1993A&A...272..153T}. NGC\,3077 sits at the edge of this extended gaseous arm. Gas might have flowed inwards to trigger the central starburst \citep{2014MNRAS.445.1694L}.  The concentration of HI, CO, and the dust in the eastern side of NGC\,3077, collectively referred to as the ``Garland", would not originate from  NGC\,3077 itself. Instead, it is more likely to have formed during the last close encounter, much like the other young systems such as Arp's Loop and Holmberg IX.

\subsection{Survival Timescale of the S-shaped Structure}
\label{subsec: timescale of S-shape}

\citet{2015ApJ...809L...1O} discovered that NGC\,3077 has a very extended stellar halo, from which a faint ``S-shaped" structure emanates to the north and south.  Here we try to estimate when this feature was formed. Using  N-body simulations, \citet{2009ApJ...698..222P} investigated how the structure of a dwarf satellite galaxy evolves during a tidal encounter with a massive host.   Interestingly, the extended shape of the NGC\,3077's halo looks almost identical to their findings (top-middle panel in their Figure 1) for a dwarf spheroidal system on a highly eccentric orbit.  This striking resemblance demonstrates that, at least qualitatively,  the extended halo of NGC\,3077 is consistent with being formed from the  tidal interaction between NGC\,3077, and M81. 
Furthermore, it also suggests the possibility of  timing  the last pericentric passage by comparing the observed stellar density profile with the simulated profiles at various snapshots in time. \citet{2009ApJ...698..222P} showed that there is a transient excess in the radial stellar density profile when a dwarf spheroidal galaxy undergoes tidal stripping. They proposed that the radius at which this transient excess occurs, the ``break radius", can be used as a clock to measure the elapsed time since the last pericentric passage. In the case of NGC\,3077, we have found in Section \ref{sec: structure} that the break occurs at $\log(\rb/\rc) \sim 0.5$, where $\rc$ is a core radius and $\rb$ is the break radius. According to the simulation of \citet{2009ApJ...698..222P}, there exists a tight relation between the last pericentric passage, $\rm{t - t_{p}}$, and the ``break radius" $\rb$, that is given by $\rb=\rm{C}{\sigma_{0}}(\rm{t-t_{p}})$, where $\rm{C}=0.55\pm0.03$ and  $\sigma_{0}$ is the stellar velocity dispersion at the galaxy center. With our measured $\log(\rb/\rc)\sim0.5$, we then obtain $\rm{t - t_{p}}=6.3~\tcr$, where $\tcr=\rc/\sigma_{0}$ is the crossing time of NGC\,3077. Therefore, the extended halo and its characteristic S-shape were formed roughly 6 crossing times ago. This value is also very close to the one of $5.7\times\tcr$, which describes the top-middle panel of Figure 1 in \citet{2009ApJ...698..222P}. 

To go further and estimate the actual elapsed time since the last pericentric passage, we need to know the central velocity dispersion of NGC\,3077. Unfortunately, we could not find a measurement of this quantity in the literature.  When $\rc$ and $\sigma_{0}$ are given in kpc and km/sec, respectively, the crossing time is given as follows: 
\begin{eqnarray}
\tcr \sim \rc[\rm{kpc}] /(\sigma_{0}[\rm{km/s}] \times 10^9 [\rm{yr}])
\label{eq:1}
\end{eqnarray}
From the density profile, we obtain $\rc=5.1\pm0.2$ kpc (see Table \ref{tbl: str}).  If we assume the velocity dispersion of NGC\,3077 is similar to that of NGC\,185, $\sigma_0=24\pm1$km/s \citep{2010ApJ...711..361G}, we obtain $\tcr=1.7\times 10^8$yr.  Thus, the last close encounter with M81 that triggered the extension of NGC\,3077 is estimated as $6.3\tcr\sim 1$Gyr ago.  While this is an approximative estimation, it might be reasonable since \citet{2017MNRAS.467..273O} calculated close encounters between M81, M82, and NGC\,3077 happened in recent few Gyrs. \citet{2019A&A...632A.122M} also showed tidal low surface brightness features survive by 0.7-4 Gyrs after major and intermediate-mass merger events.

\subsection{Examining the \citet{2022ApJ...937L...3B} Ultra-Faint Dwarf Candidates}
\label{subsec:ufds}

\begin{figure}
\begin{center}
\includegraphics[width=240pt]{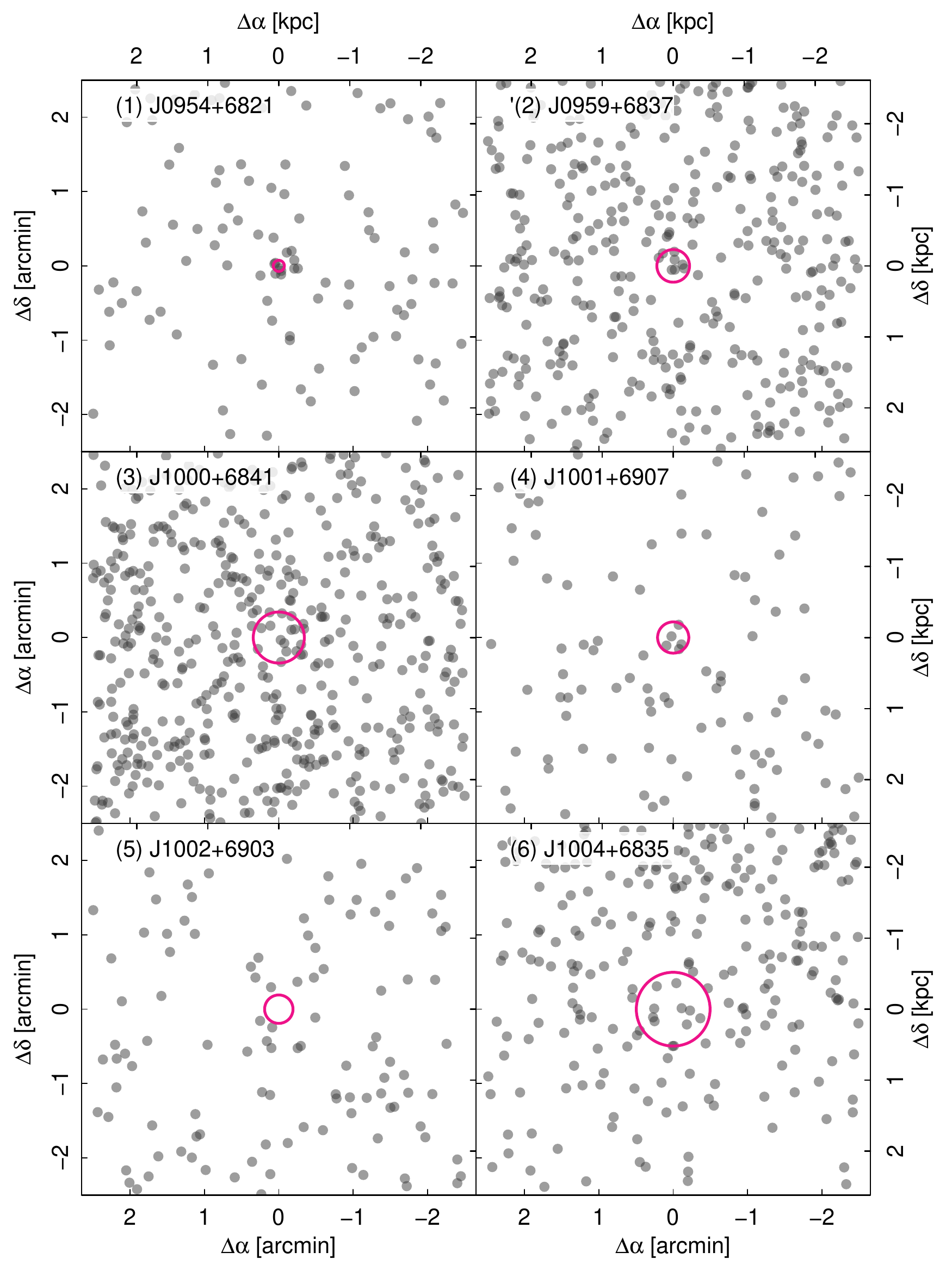}
\caption{The spatial distributions of RGB stars around the six UFD candidates reported by \citet{2022ApJ...937L...3B}. The solid lines represent the half-light radii $\rh$ as reported by those authors. The panel number corresponds to the digit in Figure \ref{fig: rgb and contour map}. } 
\label{fig: maps of ufds}
\end{center}
\end{figure}
Recently, \citet{2022ApJ...937L...3B} conducted a search for ultra-faint dwarf (UFD) candidates in the M81 Group, finding six of them. Remarkably, all of these candidates lie in close proximity to NGC\,3077 which is the smallest system of the interacting triplet. Three of these sources, J0959+6837, J1000+6841, and J1004+6835, are located well within NGC\,3077's halo structure (see black star symbols in the right panel of Figure \ref{fig: rgb and contour map}), while J1001+6907 and J1002+6903 lie around NGC3077's tidal radius (see white stars in Figure \ref{fig: rgb and contour map}).
For these five systems, \citet{2022ApJ...937L...3B} base their search on an independent dataset that they have obtained with HSC. 

We revisit the evidence for these sources using the dataset we have presented here. 
Figure \ref{fig: maps of ufds} shows the spatial distribution of RGB stars around the UFD candidates, including the sixth one, J0954+6821, defined as a definite dwarf satellite located to the south of M81 by \citet{2022ApJ...937L...3B}. This system lies within our HSC M81 Group coverage but outwith the immediate area of NGC\,3077 analyzed in this paper. The panel number (2) to (6) corresponds to the digit in Figure \ref{fig: rgb and contour map}.  While there are a few RGB stars located within the half-light radii in all cases except panel (5), we could not see any obvious over-densities in panels (2) to (6) associated with these systems.  For J0954+6821 (panel (1)), we confirm that there are some point sources of $\oi>24$ that coincide with a region of  unresolved diffuse light in our pixel frame.  From the ground-based imagery only, it is rather difficult to determine if these are RGB stars at the M81 distance of 3.6 Mpc or if they are more distant red supergiant (RSG) stars.  If these sources are intrinsically bright RSGs ($\mv\sim-6$), the distance of J0954+6821 is estimated as D $> 10$ Mpc, well behind the M81 Group. For further confirmation, deep high-resolution images from HST or JWST are required.

\section{Summary} \label{sec: summary}
We have presented the structure and metallicity distribution function of NGC\,3077.  Based on the M81 galaxy group survey with Hyper Suprime-Cam on the Subaru Telescope, we have constructed the deep CMD that reached the old RGB population of NGC\,3077 and covered well beyond the tidal radius, which allows us to derive the structural properties and stellar content of the peripheral regions.  NGC\,3077 has the extended stellar halo and the ``S-shaped" tidal feature, which is not well-matched by the position angle and the ellipticity derived from the fit to the full star count distribution. The average metallicity of individual RGBs decreases with the distance from the NGC\,3077 center. The metallicity at the S-shape area shows a comparable or slightly metal-poor population with that of $r\sim4\times\roh$ area, implying this structure being made up of stars that have been stripped from the outermost part of NGC 3077 due to the tidal interaction with M81 and M82. The old stellar component of NGC\,3077 is on the mass-metallicity relation of nearby low-mass ellipticals and shows the mild metallicity gradient $\gmh=-0.14$ dex $\roh^{-1}$, which is comparable with that of NGC\,185. These results, as well as the oxygen abundance of NGC\,3077, and M81, suggest that this peculiar galaxy had been a normal dwarf elliptical galaxy before the interaction. The close encounters with M81 and M82 might supply the gas to NGC\,3077 center and induce the current central starburst.  The ongoing strong tidal effect on NGC\,3077 is also stripping the stellar constituent from the outer envelope, which is now seen as the ``S-shaped" structure. 

\begin{acknowledgments}
We are grateful to the entire staff at Subaru Telescope and the HSC team. This research is based on data collected at the Subaru Telescope, which is operated by the National Astronomical Observatory of Japan. We are honored and grateful for the opportunity of observing the Universe from Maunakea, which has the cultural, historical, and natural significance in Hawaii.
This paper makes use of software developed for Vera C. Rubin Observatory. We thank the Rubin Observatory for making their code available as free software at http://pipelines.lsst.io/.  The Pan-STARRS1 Surveys (PS1) have been made possible through contributions of the Institute for Astronomy, the University of Hawaii, the Pan-STARRS Project Office, the Max-Planck Society and its participating institutes, the Max Planck Institute for Astronomy, Heidelberg and the Max Planck Institute for Extraterrestrial Physics, Garching, The Johns Hopkins University, Durham University, the University of Edinburgh, Queen’s University Belfast, the Harvard-Smithsonian Center for Astrophysics, the Las Cumbres Observatory Global Telescope Network Incorporated, the National Central University of Taiwan, the Space Telescope Science Institute, the National Aeronautics and Space Administration under Grant No. NNX08AR22G issued through the Planetary Science Division of the NASA Science Mission Directorate, the National Science Foundation under Grant No. AST-1238877, the University of Maryland, and Eotvos Lorand University (ELTE) and the Los Alamos National Laboratory.
S.O. acknowledges support in part from JSPS Grant-in-Aid for Scientific Research (18H05875, 20K04031, 20H05855).  N.A. thanks the Brain Pool program for financial support, which is funded by the Ministry of Science and ICT through the National Research Foundation of Korea (2018H1D3A2000902).
For the purpose of open access, the author has applied a Creative Commons Attribution (CC BY) licence to any Author Accepted Manuscript version arising from this submission.

\end{acknowledgments}


\bibliography{ms}{}
\bibliographystyle{aasjournal}



\end{document}